\PassOptionsToPackage{hidelinks}{hyperref}
\documentclass[pdflatex,sn-mathphys-ay]{sn-jnl}

\usepackage{xurl}

\usepackage{amsmath,amssymb}
\usepackage{graphicx}
\usepackage{xcolor}
\usepackage[percent]{overpic}
\usepackage{eso-pic}
\newcommand{\maybeoverpic}[3][]{%
  \IfFileExists{#2}{%
    \begin{overpic}[#1]{#2}%
      #3%
    \end{overpic}%
  }{%
    \fbox{\parbox[c][0.25\textwidth][c]{0.9\linewidth}{\centering Missing file:\\\texttt{\detokenize{#2}}}}%
  }%
}

\usepackage{grffile}
\usepackage{subcaption}
\usepackage{siunitx}
\usepackage{booktabs}

\newcommand{\maybeincludegraphics}[2][]{%
  \IfFileExists{#2}{\includegraphics[#1]{#2}}{%
    \fbox{\parbox[c][3cm][c]{0.95\linewidth}{\centering Missing file:\\\texttt{\detokenize{#2}}}}%
  }%
}

\begin{document}

\AddToShipoutPictureBG*{
  \AtPageLowerLeft{
    \put(70, 40){
      \parbox{\textwidth}{
        \footnotesize \textbf{Note:} This is a pre-print of an article published in \textit{Experimental Astronomy}.\\
        The final authenticated version is available online at: \url{https://doi.org/10.1007/s10686-026-10063-x}
      }
    }
  }
}
\title[SORAMAME]{Development of a cosmic ray detector using CMOS sensors embedded in smartphones and Raspberry Pi devices\thanks{This is a pre-print of an article published in \textit{Experimental Astronomy}. The final authenticated version is available online at: \url{https://doi.org/10.1007/s10686-026-10063-x}}}

\author*[1]{\fnm{Wakiko} \sur{Takano}}\email{r201970105fg@jindai.jp}
\author[1,2]{\fnm{Shigeharu} \sur{Udo}}
\author[3]{\fnm{Atsushi} \sur{Shiomi}}
\author[1,2]{\fnm{Kinya} \sur{Hibino}}

\affil*[1]{\orgname{Research Institute for Engineering, Kanagawa University}, \orgaddress{\city{Yokohama}, \postcode{221-8686}, \country{Japan}}}
\affil[2]{\orgname{Faculty of Engineering, Kanagawa University}, \orgaddress{\city{Yokohama}, \postcode{221-8686}, \country{Japan}}}
\affil[3]{\orgname{College of Industrial Technology, Nihon University}, \orgaddress{\city{Narashino}, \postcode{275-8575}, \country{Japan}}}

\abstract{
Cosmic rays are ubiquitous; however, their direct observation traditionally demands specialized, high-cost hardware and significant technical expertise, presenting a high barrier for non-specialist environments such as schools and community settings. We present \textbf{SORAMAME}, a smartphone and tablet application that lowers this barrier by repurposing built-in CMOS image sensors as particle detectors. The system enables real-time recording and visualization of particle-like events without additional hardware, integrating on-device extraction---calibration, noise filtering, and track-candidate detection---with cloud-based data management. 

By simplifying the detection process, SORAMAME facilitates widespread adoption across diverse user groups, fostering an environment where educational outreach can transition into large-scale data collection. This scalability is particularly significant given the unprecedented number of internet-connected consumer devices equipped with silicon CMOS image sensors. Despite the inherent constraints of consumer-grade sensors, our in-flight validation and Raspberry Pi-based measurements successfully captured altitude and latitude-dependent variations in particle flux consistent with geomagnetic shielding. These results suggest that lowering barriers to participation in observation not only serves educational purposes but also has the potential to contribute to future scientific breakthroughs through the development of global citizen science.
}

\maketitle

\section{Introduction}
Cosmic rays---high-energy particles originating from space---continuously bombard the Earth.
Secondary particles produced in the atmosphere reach the ground and can be detected as stochastic
events. While cosmic-ray science has a long history, educational access is often limited by the cost,
size, and operational complexity of traditional detectors. In parallel, modern society is witnessing the rapid expansion of the IoT ecosystem, with the number of internet-connected devices projected to reach 22 billion by 2026 and exceed 40 billion by 2034 \citep{statista2024}.
These sensors are designed for imaging light, but they also respond to charged particles by
generating electron--hole pairs along the particle path in silicon, making CMOS sensors an attractive
platform for low-cost and widely deployable particle detection and science communication.


Several citizen-science efforts have demonstrated the feasibility of using consumer electronics to record particle-like events. CRAYFIS explored the concept of a large-scale distributed detector network using smartphones and studied the response of phone CMOS sensors to ionizing particles \citep{whiteson2014}. DECO used smartphones as pocket-sized particle detectors and reported particle-like candidates observed with mobile cameras under light-shielded conditions \citep{vandenbroucke2015}. CREDO further extended the distributed-observatory concept to heterogeneous detectors, including consumer devices, and targeted large-scale correlations in cosmic-ray-related events \citep{homola2018}. 
Related studies have also examined CMOS-sensor responses for radiation detection and classification \citep{cogliati2014}, and we previously reported early educational and outreach use cases for consumer CMOS sensor as particle detectors \citep{takano2023}\citep{takano2025_pos}. Among these reports, we introduced SORAMAME, a smartphone-based application for participatory astroparticle observation. ``SORAMAME'' is a Japanese word meaning broad bean and also stands for Scientific Observation and Research on Astroparticles using Mobile and Accessible MEthods. The platform was developed with explicit emphasis on (i) real-time visualization for engagement, (ii) usability in classrooms without external hardware, and (iii) data sharing and export for collaborative inquiry. In this paper, we employ SORAMAME for quantitative validation in aircraft environments by demonstrating an altitude-dependent increase in the detection rate on two flights, and we further extend the platform to a compact embedded detector to probe route-to-route variations related to geomagnetic shielding.

This paper is organized as follows. Section~2 describes the detection principle of CMOS image
sensors. Section~3 summarizes the SORAMAME system and data workflow. Section~4 presents the
flight experiments and embedded-detector measurements. Section~5 discusses the main findings and
interpretation. Section~6 summarizes limitations and future work. 


\section{Principle: CMOS image sensors as particle detectors}
\begin{figure}[t]
\centering
\maybeincludegraphics[width=0.8\linewidth]{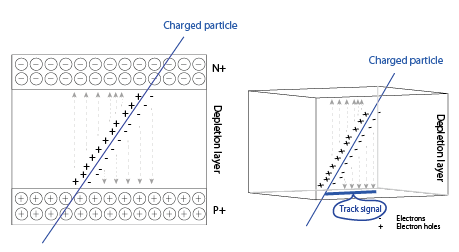}
\caption{Cross-sectional (left) and simplified 3D (right) schematics of charge generation and drift in a silicon sensor. A depletion region is formed at the p--n junction, and when a charged particle traverses this region it creates electron--hole pairs by ionization. The built-in electric field separates the carriers, driving electrons toward the p-type side and holes toward the n-type side, where they are collected. The right panel also illustrates how the particle path length within the depletion region is projected onto the sensor plane, determining the apparent track length in the pixelated image.}

\label{fig:depletion_layer}
\end{figure}

\begin{figure}[t]
  \centering
  \begin{minipage}[t]{0.65\linewidth}
    \centering
    \maybeincludegraphics[width=\linewidth]{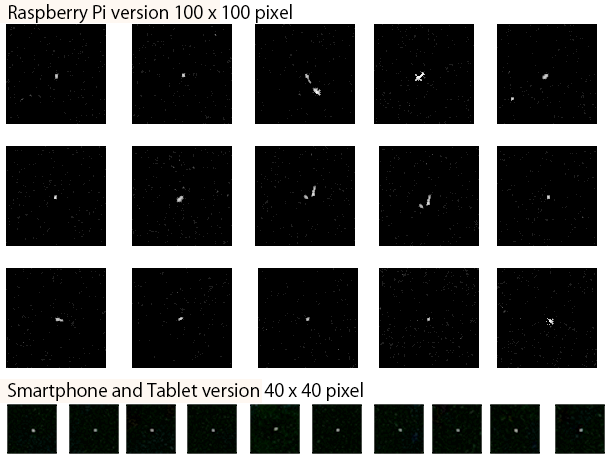}
  \end{minipage}
  \hfill
  \begin{minipage}[t]{0.33\linewidth}
    \centering
    \maybeincludegraphics[width=\linewidth]{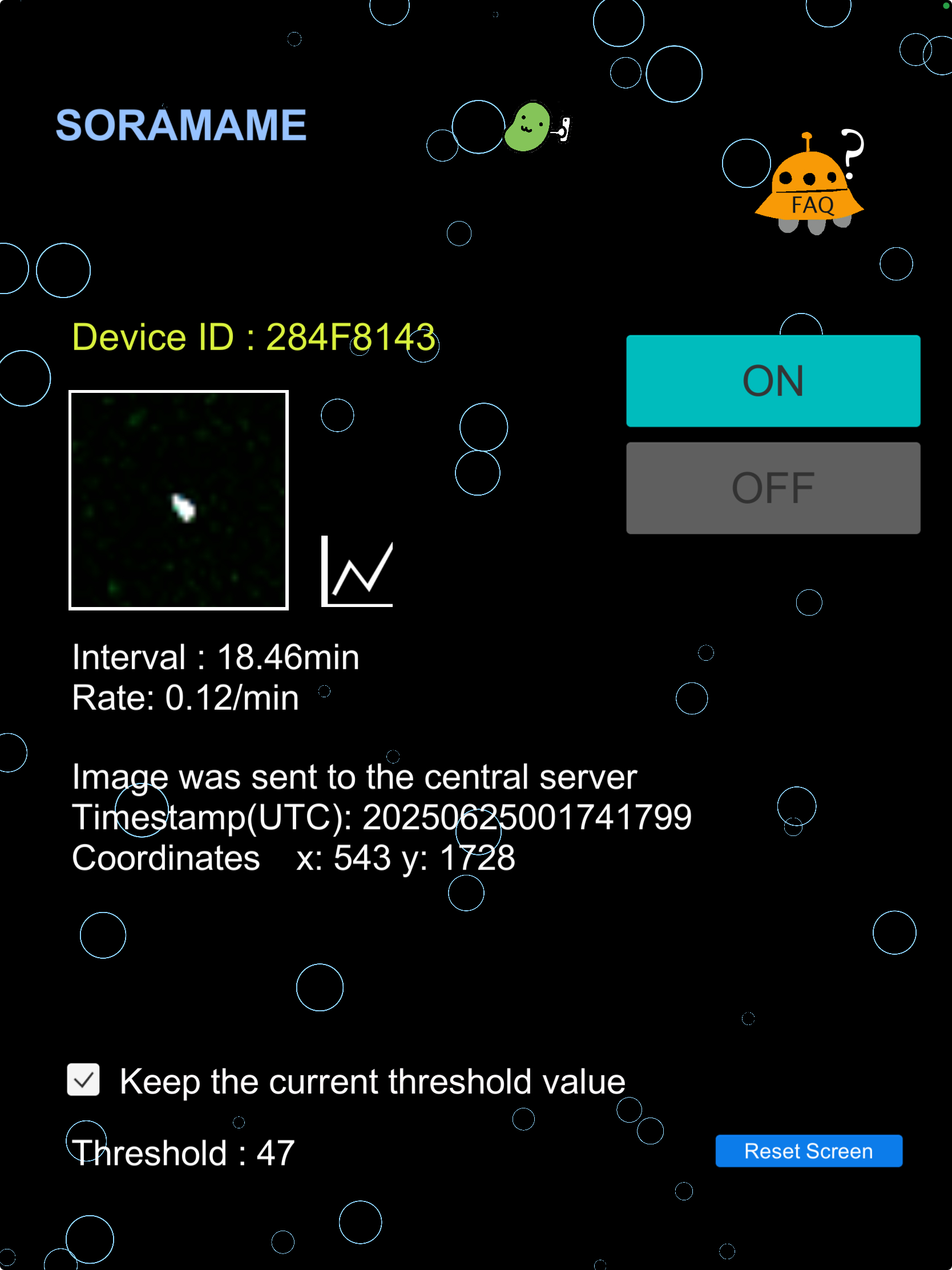}
  \end{minipage}
\caption{Examples of charged-particle candidates detected with consumer CMOS sensors (left) and the SORAMAME app (right). The left-panel images are shown with adjusted display levels (brightness/contrast) for visualization only. Most candidates are low-contrast, faint dots or short streaks, while occasional high-deposit events appear brighter.}

  \label{fig:tracks_and_ui}
\end{figure}

A CMOS image sensor consists of an array of pixels, each including
a photodiode and readout circuitry. When ionizing radiation or charged particles traverse the silicon, they deposit energy and create electron-hole pairs, primarily within the pixel depletion region, where the internal electric field sweeps the carriers toward the electrodes, as illustrated schematically in Figure~\ref{fig:depletion_layer}.
After amplification and analog-to-digital conversion, events are recorded on
otherwise dark frames as localized clusters or short streaks. In practice,
many events are only slightly above the noise floor and appear as faint,
low-contrast dots or thin lines, while particles that stop in the sensitive
volume or deposit a large amount of energy can produce brighter clusters or tracks; representative candidates are shown in Figure~\ref{fig:tracks_and_ui}. 

\section{System overview of the SORAMAME app}

\subsection{On-device processing}
SORAMAME runs a lightweight on-device process that identifies particle-like candidates in successive camera frames and records compact summaries (e.g., timestamp, pixel coordinates, and simple morphology/brightness features). This design enables continuous operation on consumer devices and supports reproducible experiments across different platforms without external readout electronics. The same high-level workflow is implemented for both the mobile-app version and the Raspberry Pi + CMOS camera version; however, the acquisition mode differs by platform. On smartphones and tablets, SORAMAME analyzes frames streamed in video mode, whereas the Raspberry Pi system operates in still-image mode with user-defined shutter speed, continuously capturing full-resolution images and performing real-time analysis.

\subsubsection{Noise-list construction and signal detection}

The smartphone/tablet version and the Raspberry Pi version employ nearly identical processing to detect charged-particle tracks. A key issue in charged-particle detection with CMOS sensors is discrimination against sensor noise. In practice, the noise can be broadly categorized into two types: (i) random noise, which appears over a wide area with relatively low intensity, and (ii) fixed-pattern noise, which repeatedly appears at specific pixel coordinates with intensities comparable to real signals. To suppress the former, the system adjusts the intensity threshold according to the current noise level. To reject the latter, the coordinates of previously detected noise clusters are stored in a noise list, and each newly detected cluster is checked against this list.

Using OpenCV, the pipeline first searches for clusters of activated pixels based on an intensity threshold tuned to the current noise level. If an excessive number of clusters is detected, the algorithm continues analyzing subsequent frames while gradually increasing the threshold. During the first few minutes of operation, the coordinates of detected clusters are stored as a noise list. After this initial period, newly detected clusters are treated as signal candidates, while the system still keeps appending entries to the noise list until it reaches 20{,}000 items. In this way, random noise is mainly controlled by threshold adjustment, whereas fixed-coordinate noise is identified through comparison with the accumulated noise list. This procedure enables both (i) an appropriate noise-level (threshold) setting and (ii) the construction of a robust noise list, which together lead to stable signal detection. In practice, completing this process requires at least several hours. Once completed, observation experiments can be performed reliably under a wide range of environmental conditions.

In typical operation, the adaptive threshold converges to values of approximately 1--50 on smartphones/tablets, depending on the device and ambient conditions. In contrast, the Raspberry Pi implementation usually stabilizes at much lower thresholds of around 2--3; under outdoor summer observations, the threshold can transiently increase to $\sim$20 due to elevated noise levels. Repeated measurements have also indicated that genuine cosmic-ray tracks are often very faint. Therefore, the system is intentionally designed to keep the threshold as low as practicable while relying on the noise-list mechanism to maintain robustness against spurious clusters.

\subsection{Visualization and user interface}
Detected particle tracks are displayed in real time on the application interface. In addition, each detected candidate is displayed as a ring at the corresponding sensor
coordinate, allowing users to see where events occur on the pixel plane (see the right panel of Figure~\ref{fig:tracks_and_ui}).
Rings gradually fade over time so that the interface reflects recent activity
without becoming saturated. This immediate visual feedback is also useful
for diagnosing residual noise, such as repeated detections at fixed positions or
edge clustering.

\subsection{Cloud upload, web dashboard, and data export}
\begin{figure}[t]
  \centering
  \includegraphics[width=0.92\linewidth]{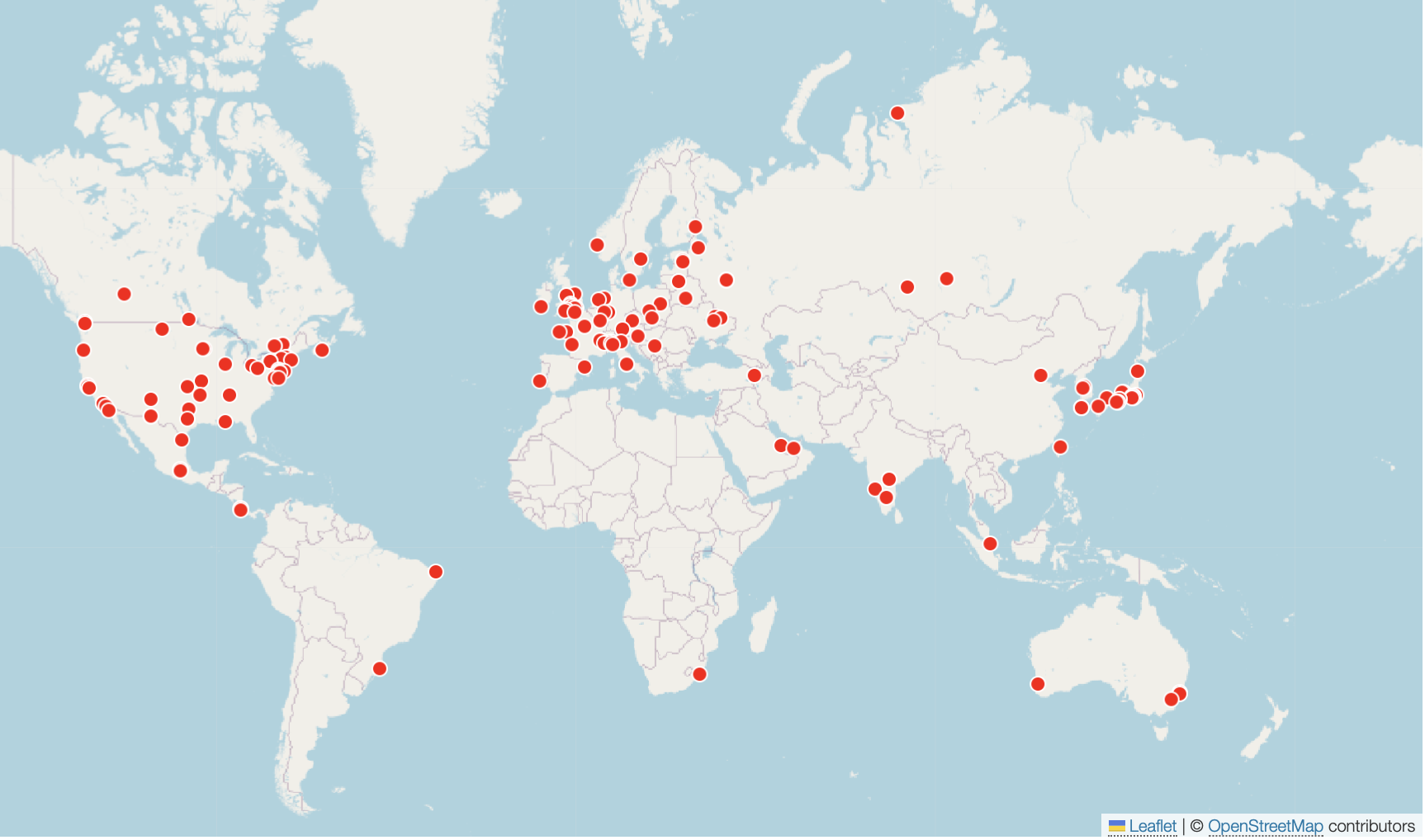}
  \caption{World map of observation locations shown on the SORAMAME web dashboard.}
  \label{fig:soramame_user_map}
\end{figure}

Detected event candidates are uploaded to a cloud server when connectivity is available.
A public web dashboard~\cite{soramame_dashboard} displays recent events, device-specific detection-rate time series,
and a world map of observation locations, shown in Figure~\ref{fig:soramame_user_map}.
Data can be downloaded in CSV format, enabling further analysis in spreadsheets or
programming environments and supporting inquiry-based learning workflows.

\section{Experiments}
\subsection{Validation 1: Altitude dependence with a smartphone (BKK--HND)}
\label{sec:exp1}

\subsubsection{Measurement setup}
This experiment was conducted on 15 March 2024 on JL32 (BKK--HND; departure 09:55 Bangkok).
We operated the SORAMAME app on an \textit{iPhone 13 Pro} using the rear main camera and recorded event counts during the flight.
The main camera uses a Sony stacked 12MP CMOS image sensor (IMX703) with a pixel pitch of 1.9~$\mu$m, and an optical format of approximately $(1/1.68)''$ corresponding to an active area of about $(7.6 \times 5.7)$~mm$^2$ ($\sim 43.5$~mm$^2$). \cite{ltec_imx703}
To prevent light leakage, we covered the camera area with opaque shielding tape.

The measurement was started after takeoff and data acquisition was continued using the aircraft’s paid in-flight internet (Wi-Fi) service.


\subsubsection{Ground vs.\ flight comparison (Bangkok$\rightarrow$Haneda)}
We compared event-candidate rates between a ground reference dataset and a flight dataset recorded on the Bangkok$\rightarrow$Haneda route. Each dataset consists of timestamped event candidates, and event counts were modeled as a Poisson process. For the flight data, we restricted the analysis to the stable-cruise portion of the trajectory, identified using the recorded altitude profile as the interval where the aircraft remained above 30{,}000~ft and the altitude was approximately steady. We then binned event counts in 20-minute intervals within this cruise-only segment.

The exposure time was defined consistently with this procedure as the duration of the selected cruise-only interval, aligned to the 20-minute bin boundaries (i.e., from the start of the first included bin to the end of the last included bin). By discarding non-cruise periods and using a bin-aligned exposure definition over a clearly defined steady-flight segment, the resulting rates provide a more reliable summary of in-flight event-candidate activity under stable operating conditions.

For the ground reference measurement, $N_g=38$ unique events were recorded over
$T_g=10.33~\mathrm{h}$, corresponding to $\hat{\lambda}_g=3.68~\mathrm{h}^{-1}$
(exact Poisson 68.27\% CI: $3.08$--$4.38~\mathrm{h}^{-1}$; Garwood).
For the flight dataset, $N_f=87$ unique events were recorded over
$T_f=4.50~\mathrm{h}$, corresponding to $\hat{\lambda}_f=19.33~\mathrm{h}^{-1}$
(exact Poisson 68.27\% CI: $17.26$--$21.64~\mathrm{h}^{-1}$; Garwood).
The incidence-rate ratio between flight and ground is $\mathrm{IRR}=5.26$
(exact conditional 68.27\% CI: $4.25$--$6.51$).
An exact conditional Poisson rate test rejects the null hypothesis of equal rates
(two-sided $p=2.2\times10^{-19}$), corresponding to a Gaussian-equivalent significance of
$Z=9.00\sigma$.

\begin{figure}[htbp]
  \centering
  \includegraphics[width=0.7\linewidth]{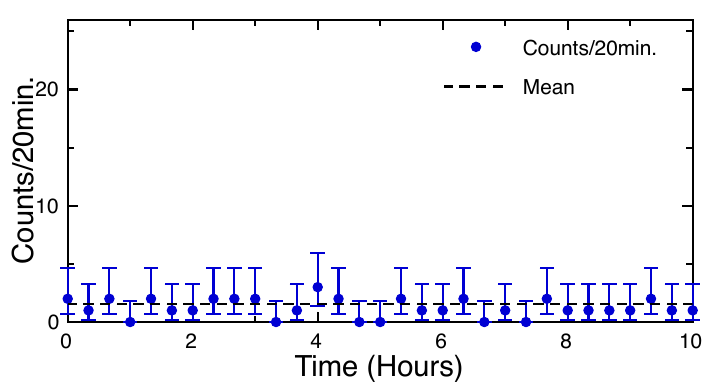}
  \caption{Ground-reference time series. Event counts are binned in 20-minute
  intervals and shown as points. Error bars indicate exact Poisson 68.27\% confidence intervals (Garwood). }
  \label{fig:ground_10min_points_bkkhnd}
\end{figure}

\begin{figure}[htbp]
  \centering
  \includegraphics[width=0.7\linewidth]{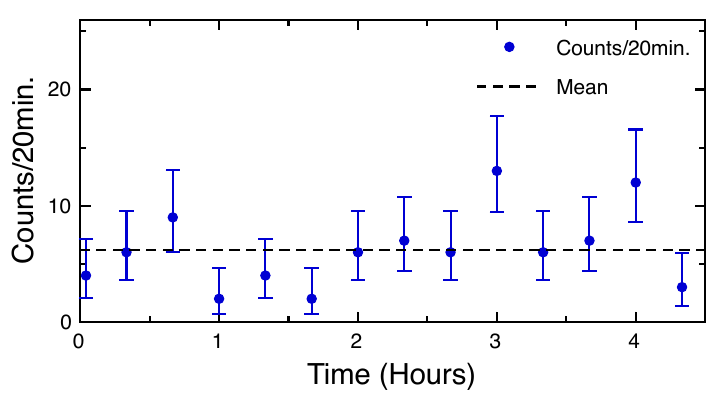}
  \caption{Flight time series for the Bangkok$\rightarrow$Haneda dataset. Event counts are binned in 20-minute
  intervals and shown as points. Error bars indicate exact Poisson 68.27\% confidence intervals (Garwood). }
  \label{fig:flight_10min_points_bkkhnd}
\end{figure}

Figure~\ref{fig:ground_10min_points_bkkhnd} and Figure~\ref{fig:flight_10min_points_bkkhnd}
show the binned time series for the ground and flight datasets, respectively.
Counts are binned in 20-minute intervals and plotted as points; error bars indicate exact Poisson 68.27\% confidence intervals (Garwood).
Table~\ref{tab:rate_comparison_ground_vs_bkkhnd} summarizes the full-interval rates using the same confidence level.

\begin{table}[t]
  \centering
  \caption{Full-interval event-candidate rates for ground and flight datasets (Bangkok$\rightarrow$Haneda), modeled as Poisson processes. Rates are reported with exact Poisson 68.27\% confidence intervals (Garwood). IRR denotes the flight/ground incidence-rate ratio with an exact conditional 68.27\% confidence interval.}
  \label{tab:rate_comparison_ground_vs_bkkhnd}
  \begin{tabular}{lrrr}
    \toprule
    Dataset & Events $N$ & Exposure $T$ [h] & Rate $\hat{\lambda}$ [$\mathrm{h}^{-1}$] (68.27\% CI) \\
    \midrule
    Ground & 38 & 10.33 & 3.68 (3.08--4.38) \\
    Flight (BKK$\rightarrow$HND) & 87 & 4.50 & 19.33 (17.26--21.64) \\
    \midrule
    \multicolumn{3}{l}{Incidence-rate ratio (flight/ground)} & $\mathrm{IRR}=5.26$ (4.25--6.51) \\
    \multicolumn{3}{l}{Exact Poisson rate test ($H_0:\lambda_f=\lambda_g$; two-sided)} & $p=2.2\times10^{-19}$ \\
    \multicolumn{3}{l}{Gaussian-equivalent significance (two-sided)} & $Z=9.00\sigma$ \\
    \bottomrule
  \end{tabular}
\end{table}


\subsection{Validation 2: Reproducibility on a second flight (YUL--NRT)}
\label{sec:exp2}

\subsubsection{Measurement setup}
As in Experiment~1, we used an iPhone 13 Pro and operated SORAMAME with the rear camera fully
light-shielded to record candidate events during the flight. This experiment was conducted on the Montreal--Narita (YUL--NRT) route.
During the flight, we also recorded the phone's location data to reconstruct the flight route,
as shown in Figure~\ref{fig:flight_map}.

Due to battery depletion, continuous recording was terminated before the end of the flight.

\subsubsection{Ground vs.\ flight comparison}
We compared event-candidate rates between a ground reference dataset in Japan and a flight dataset recorded on the Montreal--Narita (YUL--NRT) route. Each dataset consists of timestamped event candidates, and event counts were modeled as a Poisson process. For the flight data, we excluded records near takeoff and restricted the analysis to the stable-cruise portion of the trajectory, identified from the recorded altitude profile as the interval where the aircraft remained above 30{,}000~ft and the altitude was approximately steady. Event counts were then binned in 20-minute intervals within this cruise-only segment.

The exposure time was defined consistently with this procedure as the duration of the selected cruise-only interval, aligned to the 20-minute bin boundaries (i.e., from the start of the first included bin to the end of the last included bin). 

For the ground reference measurement, $N_g=38$ unique events were recorded over
$T_g=10.33~\mathrm{h}$, corresponding to $\hat{\lambda}_g=3.68~\mathrm{h}^{-1}$
(exact Poisson 68.27\% CI: $3.08$--$4.38~\mathrm{h}^{-1}$; Garwood).
For the flight dataset, $N_f=100$ unique events were recorded over
$T_f=3.67~\mathrm{h}$, corresponding to $\hat{\lambda}_f=27.27~\mathrm{h}^{-1}$
(exact Poisson 68.27\% CI: $24.53$--$30.25~\mathrm{h}^{-1}$; Garwood).
The incidence-rate ratio between flight and ground is $\mathrm{IRR}=7.42$
(exact conditional 68.27\% CI: $6.02$--$9.13$).
An exact conditional Poisson rate test rejects the null hypothesis of equal rates
(two-sided $p=1.0\times10^{-29}$), corresponding to a Gaussian-equivalent significance of
$Z=11.32\sigma$, indicating a statistically significant difference in the event-candidate rate
between the flight and ground measurements.

\begin{figure}[htbp]
  \centering
  \includegraphics[width=0.7\linewidth]{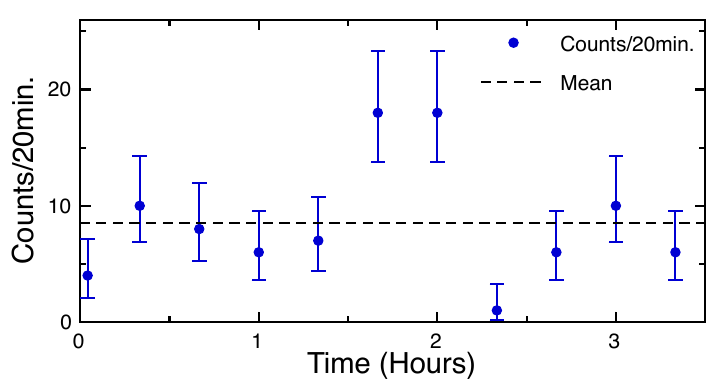}
  \caption{Flight time series for the YUL--NRT dataset. Event counts are binned in 20-minute
  intervals and shown as points. Error bars indicate exact Poisson 68.27\% confidence intervals (Garwood). }
  \label{fig:flight_10min_points_yulnrt}
\end{figure}

\begin{figure}[t]
  \centering
  \maybeincludegraphics[width=0.60\linewidth]{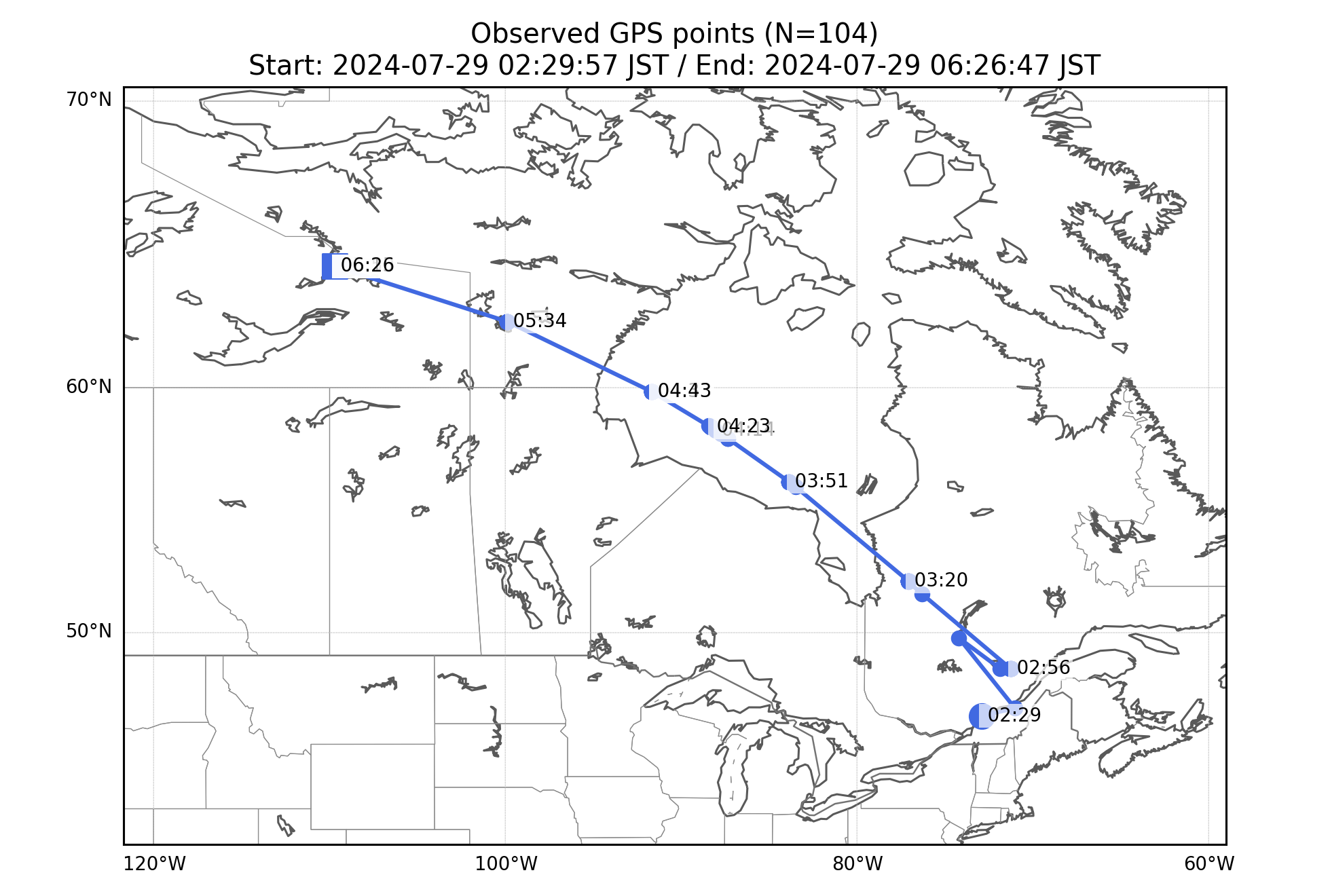}
  \caption{Flight route for Experiment~2. The path is reconstructed from the spatio-temporal data
  (latitude, longitude, and time) captured by the onboard iPhone 13 Pro. (Map data by OpenStreetMap)}
  \label{fig:flight_map}
\end{figure}

Figure~\ref{fig:ground_10min_points_bkkhnd} and Figure~\ref{fig:flight_10min_points_yulnrt}
show the binned time series for the ground and flight datasets, respectively.
Counts are binned in 20-minute intervals and shown as points; error bars indicate exact Poisson 68.27\% confidence intervals (Garwood).
Table~\ref{tab:rate_comparison_ground_vs_yulnrt} summarizes the corresponding full-interval event rates and rate comparison using the same confidence level.

\begin{table}[t]
  \centering
  \caption{Full-interval event-candidate rates for ground and flight datasets (YUL--NRT), modeled as Poisson processes. Values are rates with exact Poisson 68.27\% confidence intervals (Garwood). IRR denotes the flight/ground incidence-rate ratio with an exact conditional 68.27\% confidence interval.}
  \label{tab:rate_comparison_ground_vs_yulnrt}
  \begin{tabular}{lrrr}
    \toprule
    Dataset & Events $N$ & Exposure $T$ [h] & Rate $\hat{\lambda}$ [$\mathrm{h}^{-1}$] (68.27\% CI) \\
    \midrule
    Ground (Japan) & 38 & 10.33 & 3.68 (3.08--4.38) \\
    Flight (YUL--NRT) & 100 & 3.67 & 27.27 (24.53--30.25) \\
    \midrule
    \multicolumn{3}{l}{Incidence-rate ratio (flight/ground)} & $\mathrm{IRR}=7.42$ (6.02--9.13) \\
    \multicolumn{3}{l}{Exact Poisson rate test ($H_0:\lambda_f=\lambda_g$; two-sided)} & $p=1.0\times10^{-29}$ \\
    \multicolumn{3}{l}{Gaussian-equivalent significance (two-sided)} & $Z=11.32\sigma$ \\
    \bottomrule
  \end{tabular}
\end{table}

\subsection{Latitude dependence and geomagnetic effects with an embedded CMOS detector}
\label{sec:exp3}
This measurement is designed to evaluate geomagnetic effects explicitly. Whereas Experiments~1 and~2 validate altitude dependence using smartphones, Experiment~3 uses an embedded CMOS detector to compare routes spanning different geomagnetic latitudes.

\subsubsection{Prototype detector}
\begin{figure}[t]
  \centering
  \includegraphics[width=0.4\linewidth]{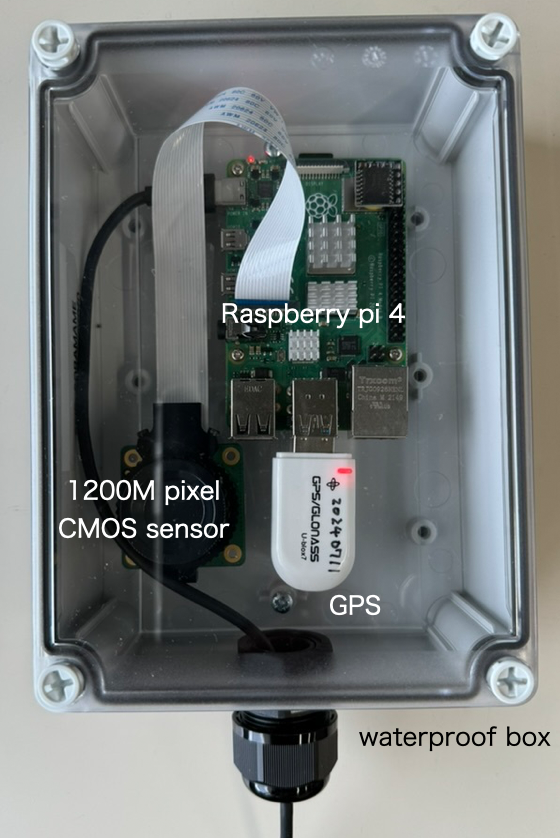}
  \caption{Prototype detector based on a Raspberry~Pi~4 and an HQ camera module enclosed in a waterproof case.}
  \label{fig:prototype_detector}
\end{figure}

We built a compact detector consisting of a Raspberry~Pi~4 and a Raspberry~Pi High Quality (HQ) Camera Module enclosed in a waterproof case, shown in Figure~\ref{fig:prototype_detector}.
The HQ camera uses a Sony IMX477R back-illuminated stacked CMOS image sensor (optical format: type~1/2.3), with 12.3~MP effective pixels (4056\(\times\)3040) and a pixel pitch of 1.55~\(\mu\)m.\cite{rpi_hq_camera,sony_imx477_flyer,distrelec_hq_camera_tds}
This corresponds to an active imaging area of approximately \(6.29 \times 4.71\)~mm\(^2\) (about 29.6~mm\(^2\)).\cite{rpi_hq_camera,sony_imx477_flyer,distrelec_hq_camera_tds}

The shutter speed (exposure time) was set to 1~s.
The system continuously captured full-resolution still images while performing real-time on-device analysis, and recorded time-stamped detections together with the corresponding signal-candidate images and associated features.
No internet connection was used during operation, and all observations were stored locally on the Raspberry Pi's microSD card for later collection.

\subsubsection{Flight campaigns}
\begin{figure}[htbp]
  \centering
  \includegraphics[width=0.95\linewidth]{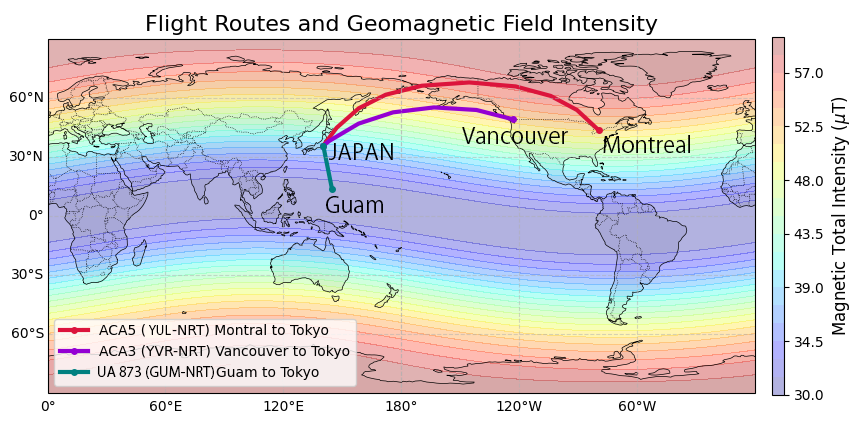}
  
\caption{Comparison of flight trajectories and a global geomagnetic-field reference map. The map illustrates the great-circle routes for ACA5 (Montreal--Narita), ACA3 (Vancouver--Narita), and the UA 873 (Guam--Narita) route, overlaid on the total magnetic intensity ($\mu$T) calculated via the IGRF-13 model~\cite{alken2021igrf13}. The magnetic intensity is shown here as a geographic reference; the relevant proxy for cosmic-ray access is the geomagnetic cutoff rigidity, which generally decreases toward higher geomagnetic latitudes. Accordingly, the trans-polar and high-latitude routes (ACA5 and ACA3) are expected to admit a larger flux of lower-rigidity primaries than the tropical Guam route, consistent with the qualitative rate ordering discussed in Section~4.3.3. (Map data by OpenStreetMap)}
\label{fig:geomagnetic_flight_map}

\end{figure}

Measurements were conducted at ground level in Japan and aboard international flights from Guam, Vancouver, and Montreal to Japan.
These routes span different geomagnetic latitudes, enabling a qualitative comparison of event rates under different geomagnetic shielding conditions, as summarized in Figure~\ref{fig:geomagnetic_flight_map}.

\subsubsection{Route-to-route comparison}

We compared event-candidate rates among three representative flight trajectories
(Montreal$\rightarrow$Japan, Vancouver$\rightarrow$Japan, and Guam$\rightarrow$Japan)
and a ground reference dataset in Japan (July 2024). Each dataset consists of
timestamped event candidates obtained by the embedded CMOS detector operated
with a fixed shutter speed of 1~s.

For the flight data, we used FlightAware-derived flight profiles to identify the
cruise portions of each trajectory and restricted the analysis to segments above
30{,}000~ft. Event counts were evaluated over these cruise-only intervals using
10-minute-bin-aligned exposure. When multiple detections shared an identical timestamp, they were consolidated so that
$N$ denotes the number of unique timestamps.



Table~\ref{tab:route_rate_summary_bin10aligned} summarizes the resulting event counts and Poisson full-interval rates with exact 68.27\% confidence intervals (Garwood).
For the ground reference dataset, $N=3076$ unique events were recorded over $T=179.83~\mathrm{h}$,
corresponding to $17.10~\mathrm{h}^{-1}$ (exact Poisson 68.27\% CI: $16.79$--$17.41~\mathrm{h}^{-1}$; Garwood).
For the flights, the estimated rates were $53.63~\mathrm{h}^{-1}$ (Guam$\rightarrow$Japan;
$N=143$, $T=2.67~\mathrm{h}$; exact Poisson 68.27\% CI: $49.15$--$58.50$),
$65.51~\mathrm{h}^{-1}$ (Vancouver$\rightarrow$Japan; $N=404$, $T=6.17~\mathrm{h}$;
exact Poisson 68.27\% CI: $62.25$--$68.93$),
and $77.49~\mathrm{h}^{-1}$ (Montreal$\rightarrow$Japan; $N=607$, $T=7.83~\mathrm{h}$;
exact Poisson 68.27\% CI: $74.35$--$80.76$).

Relative to the ground reference, the incidence-rate ratios (IRR) were
$\mathrm{IRR}=3.14$ (Guam$\rightarrow$Japan; exact conditional 68.27\% CI: 2.87--3.43; two-sided $p=4.8\times10^{-30}$; $Z=11.39\sigma$),
$\mathrm{IRR}=3.83$ (Vancouver$\rightarrow$Japan; exact conditional 68.27\% CI: 3.63--4.04; two-sided $p=2.4\times10^{-102}$; $Z=21.48\sigma$),
and $\mathrm{IRR}=4.53$ (Montreal$\rightarrow$Japan; exact conditional 68.27\% CI: 4.33--4.74; two-sided $p=9.1\times10^{-181}$; $Z=28.67\sigma$),
where each $p$-value is from an exact conditional Poisson rate test.

Among the flight routes, the Montreal$\rightarrow$Japan rate remained higher than
Vancouver$\rightarrow$Japan (IRR $=1.18$, exact conditional 68.27\% CI: 1.11--1.26, two-sided $p=9.4\times10^{-3}$, $Z=2.60\sigma$)
and Guam$\rightarrow$Japan (IRR $=1.45$, exact conditional 68.27\% CI: 1.31--1.59, two-sided $p=4.6\times10^{-5}$, $Z=4.08\sigma$),
whereas the difference between Vancouver$\rightarrow$Japan and Guam$\rightarrow$Japan was modest
(IRR $=1.22$, exact conditional 68.27\% CI: 1.10--1.35, two-sided $p=4.0\times10^{-2}$, $Z=2.05\sigma$).
These pairwise tests are reported as uncorrected (exploratory) comparisons. Nevertheless, the point estimates exhibit a monotonic trend consistent with geomagnetic shielding expectations: the higher-geomagnetic-latitude route (Montreal$\rightarrow$Japan) shows the highest rate, followed by Vancouver$\rightarrow$Japan and the low-latitude Guam$\rightarrow$Japan route.



\begin{figure}[htbp]
  \centering
  \includegraphics[width=0.78\linewidth]{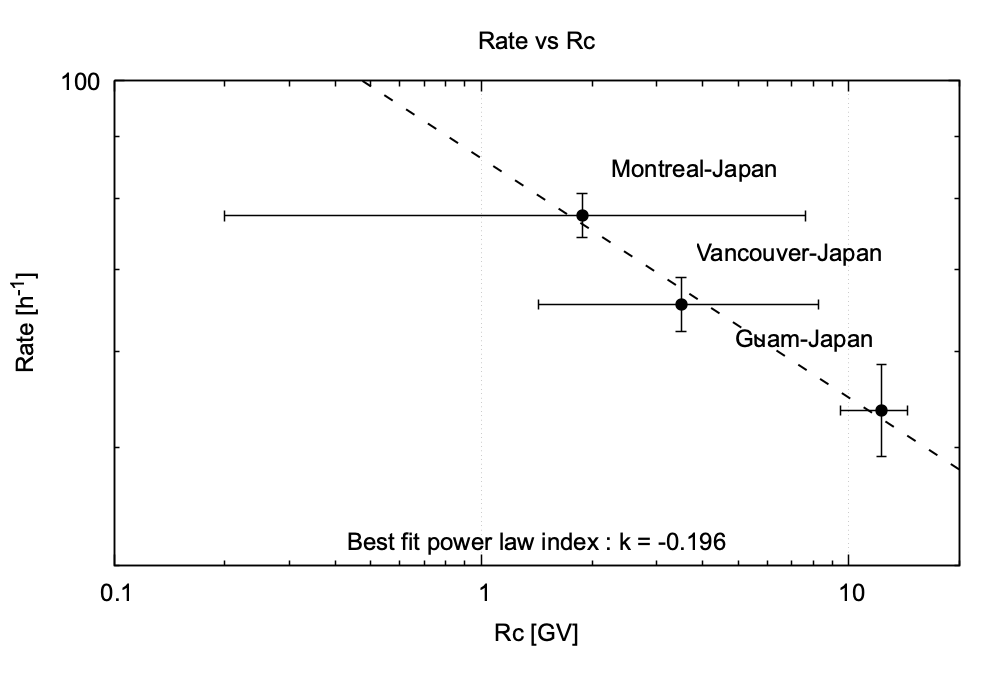}
\caption{Event rates as a function of mean cutoff rigidity for the three flight routes, showing the rigidity dependence of the observed rates. Horizontal error bars denote the 5--95\% range of cutoff rigidity along each route, and vertical error bars denote the exact 68.27\% confidence intervals (Garwood) for the measured event rates. The dashed line indicates the best-fit power-law model, $\mathrm{rate}\propto R^{k}$, with fitted index $k=-0.196$.}
    \label{fig:routes_rate_comparison_bin10aligned}
    
\end{figure}

Figure~\ref{fig:routes_rate_comparison_bin10aligned} compares the full-interval event-candidate rates as a function of the corresponding mean geomagnetic cutoff rigidity $R_c$ along each route.
Across the three routes, the rate decreases with increasing $R_c$, and a simple power-law model,
$\mathrm{Rate} \propto R_c^{k}$, provides a descriptive fit with $k \approx -0.196$.
This behavior is qualitatively consistent with geomagnetic shielding, which suppresses low-rigidity primaries at higher cutoff rigidities.
Given the limited number of routes, no quantitative interpretation of $k$ is attempted.

Figure~\ref{fig:routes_rate_comparison_bin10aligned} compares the full-interval event-candidate rates as a function of the corresponding mean geomagnetic cutoff rigidity $R_c$ along each route.
Across the three routes, the rate decreases with increasing $R_c$, and a simple power-law model,
$\mathrm{Rate} \propto R_c^{k}$, provides a descriptive fit with $k \approx -0.196$.

\begin{table}[htbp]
\centering
\caption{Event rates for the ground reference and three flight routes using 10-minute-bin-aligned exposure over FlightAware-selected cruise segments above 30{,}000~ft. Rates are modeled as Poisson processes and reported with exact Poisson 68.27\% confidence intervals (Garwood).}
\label{tab:route_rate_summary_bin10aligned}
\begin{tabular}{lrrrr}
\hline
Dataset & $N$ (unique) & $T$ [h] & Rate [$\mathrm{h}^{-1}$] & 68.27\% CI [$\mathrm{h}^{-1}$] \\
\hline
Ground (Japan, Jul 2024) & 3076 & 179.83 & 17.10 & 16.79--17.41 \\
Guam$\rightarrow$Japan & 143 & 2.67 & 53.63 & 49.15--58.50 \\
Vancouver$\rightarrow$Japan & 404 & 6.17 & 65.51 & 62.25--68.93 \\
Montreal$\rightarrow$Japan & 607 & 7.83 & 77.49 & 74.35--80.76 \\
\hline
\end{tabular}
\end{table}

\subsection{Data Visualization and Software}
The code used to generate Figures~7 and~9 was developed with the assistance of ChatGPT 5.2 and Gemini 3. All AI-assisted code was manually reviewed, verified, and executed by the authors.

\section{Discussion}
Validations~1 and~2 were designed to test whether the smartphone-based detector responds to the expected increase in secondary-particle flux at aviation altitudes and to assess reproducibility across different flights. In both measurements, the event-candidate rate remained elevated during the flight compared with ground-reference periods. This behavior is consistent with the well-known altitude dependence of the secondary cosmic-ray flux and supports the use of consumer CMOS sensors as a practical platform for qualitative demonstrations and outreach.

Importantly, no increase in thermal-noise-related event rates was observed during the cruise phase, suggesting that the elevated rates are not artifacts of sensor heating.

The Raspberry~Pi configuration used in the third measurement enables more controlled acquisition. Although the present route-to-route comparisons remain exploratory, the measured rates exhibited a latitude-ordered tendency, with higher rates on higher-geomagnetic-latitude routes. This qualitative pattern is consistent with the geomagnetic-cutoff picture, in which lower latitudes experience stronger geomagnetic shielding and therefore admit fewer low-rigidity primaries, leading to reduced secondary-particle rates at aircraft altitudes.

These results motivate future measurements with increased exposure and standardized analysis windows (e.g., stable cruise segments), enabling quantitative comparisons against geomagnetic cutoff rigidity models. In addition, further analysis of data including the ascent and descent phases will be valuable to isolate altitude-dependent effects more precisely.

\section{Limitations and future work}
Detection performance depends on device-specific noise characteristics, camera hardware, shielding quality, and operating conditions. Practically, the generation of a robust noise list necessitates several hours; therefore, for classroom applications, we recommend initiating the setup at least one day in advance. Additionally, given the substantial power consumption during continuous observation, the use of an external power source is strongly advised. Future work is required to strengthen event classification, establish cross-device calibration procedures, and develop robust uncertainty estimates that account for potential non-Poisson effects and operational variability (e.g., interruptions, orientation changes, and varying altitude profiles). On the outreach side, we plan to expand teaching materials and guided exercises, host public demonstrations, and provide analysis notebooks and curriculum-aligned modules to support inquiry-based learning.


\section{Conclusion}
We have presented SORAMAME, a portable cosmic-ray and ionizing-radiation detection system that repurposes consumer CMOS image sensors as accessible particle detectors on both smartphones and a Raspberry~Pi camera platform. Flight validations on two commercial routes (BKK--HND and YUL--NRT) show clear increases in event-candidate rates aloft relative to ground reference periods. Using a Raspberry Pi--based prototype detector with a CMOS image sensor, we further observed a latitude-ordered tendency in the route-to-route rates that is consistent with geomagnetic-cutoff variations, suggesting the expected latitude dependence.

These results indicate that even consumer-grade CMOS image sensors can capture changes in charged-particle density associated with cosmic-ray air showers under realistic operating conditions. Given the widespread availability of devices equipped with such sensors, the present study suggests a practical pathway by which simple and accessible detection tools may broaden participation in observation-based activities. In this sense, SORAMAME provides not only a reproducible platform for flight-based demonstrations, education, and outreach, but also a basis for future citizen-science efforts that may, with further validation and scaling, contribute to scientifically useful measurements.



\section*{Declarations}

\subsection*{Funding}
This work was supported by the Kitano Foundation of Lifelong Integrated Education.

\subsection*{Competing interests}
The authors declare that they have no competing interests.

\subsection*{Data availability}
The data supporting the findings of this study are available from the corresponding author upon reasonable request.

\subsection*{Code availability}
The source code used for the smartphone application and the Raspberry Pi camera analysis is publicly available at: \url{https://github.com/soramame-cosmicray}.

\subsection*{Authors' contributions}
Wakiko Takano made a major contribution to all aspects of this study, including conceptualization, methodology, formal analysis, software development, validation, visualization, investigation, data curation, project administration, funding acquisition, resource preparation, and writing of the original draft.

Kinya Hibino contributed broadly to the study through conceptualization, methodology, formal analysis, software, validation, visualization, supervision, project administration, funding acquisition, resource preparation, and review and editing of the manuscript.

Shigeharu Udo contributed to data curation, investigation, and review and editing of the manuscript.

Atsushi Shiomi contributed to data curation, investigation, and review and editing of the manuscript.

All authors read and approved the final manuscript.


\end{document}